\documentclass[12pt,a4paper,english,superscriptaddress,aps,nofootinbib]{revtex4}
\usepackage{amsmath,amssymb,graphicx}

\makeatletter
\usepackage{babel}
\usepackage[active]{srcltx}
\usepackage{graphicx,color}
\usepackage{changebar}
\usepackage{hyperref}
\usepackage[utf8]{inputenc}
\usepackage[T1]{fontenc}
\usepackage{ae}
\usepackage[caption = false]{subfig}

\usepackage{graphicx}
\usepackage{amsfonts}
\newcommand{\be}{\begin{equation}}
	\newcommand{\ee}{\end{equation}}
\newcommand{\bea}{\begin{eqnarray}}
	\newcommand{\eea}{\end{eqnarray}}

\begin{document}
	\title{Phase transitions in the logarithmic Maxwell O(3)-sigma model}
	
\author{F. C. E. Lima}
\email[]{E-mail: cleiton.estevao@fisica.ufc.br}
\affiliation{Universidade Federal do Cear\'{a} (UFC), Departamento do F\'{i}sica - Campus do Pici, Fortaleza, CE, C. P. 6030, 60455-760, Brazil.}

\author{C. A. S. Almeida}
\email[]{E-mail: carlos@fisica.ufc.br}
\affiliation{Universidade Federal do Cear\'{a} (UFC), Departamento do F\'{i}sica - Campus do Pici, Fortaleza, CE, C. P. 6030, 60455-760, Brazil.}

\begin{abstract}
We investigate the presence of topological structures and multiple phase transitions in the O(3)-sigma model with the gauge field governed by Maxwell's term and subject to a so-called Gausson's self-dual potential. To carry out this study, it is numerically shown that this model supports topological solutions in 3-dimensional spacetime. In fact, to obtain the topological solutions, we assume a spherically symmetrical ansatz to find the solutions, as well as some physical behaviors of the vortex, as energy and magnetic field. It is presented a planar view of the magnetic field as an interesting configuration of a ring-like profile. To calculate the differential configurational complexity (DCC) of structures, the spatial energy density of the vortex is used. In fact, the DCC is important because it provides us with information about the possible phase transitions associated with the structures located in the Maxwell-Gausson model in 3D.  Finally, we note from the DCC profile an infinite set of kink-like solutions associated with the parameter that controls the vacuum expectation value.

\vspace{0.2cm}
\hspace{-0.5cm}\textbf{Keyword:} Gauge field theories, Nonlinear sigma model, Configurational entropy, Extended classical solutions.
\end{abstract}

\maketitle
\newpage
\thispagestyle{empty}

\section{Introduction}

The interest in the $O(3)$-sigma model arises from the description of cosmic strings \cite{Nielsen,Tong,Hanany} as well as phenomena in condensed matter physics \cite{Kleinert}. In the low energy regime, Wilczek and Zee \cite{Wilczek} proposed that solitons of the $O(3)$-sigma model in certain conditions acquire a new statistical behavior, i. e., the possibility of fractional spin solitons \cite{Wilczek}. In 1983, Haldane \cite{Haldane} observed that the dynamics of these structures in the $O(3)$ - sigma model allow to describe the behavior of Heisenberg antiferromagnetic materials \cite{Haldane1}. The study of this model has aroused the interest of several researchers due to these and other applications in different areas of physics \cite{Schroers,MOTRUNICH,Singh}.

In 1995, Schroers showed the presence of Bogomol'nyi solitons in a gauged $O(3)$-sigma model \cite{Schroers}. He showed that the scale invariance of the sigma model can be broken by gauging a subgroup $U(1)$ of the $O(3)$ symmetry, e. g., including the Maxwell gauge field. In 1996, a new proposal was made to find topological and non-topological solitons in a model involving the Chern-Simons field \cite{Ghosh}. Subsequently, further studies on the $O(3)$-sigma model with Maxwell field were carried out, e. g., studies of the sigma model with nonminimal coupling and with Maxwell-Chern-Simons fields \cite{Cavalcante}, as well as studies of topological multivortex solutions in the Maxwell-Chern-Simons self-dual model \cite{Han}.

In a general context, the pioneering work of Nilsen and Olesen \cite{Nielsen} demonstrated the existence of static vortices in a Maxwell-Higgs theory. Later, motivated by this work, several researchers have intensively studied Maxwell's vortices in different models \cite{Lee,Atmaja,Casana3}.

Topological solutions of the gauged $O(3)$-sigma model have been extensively studied \cite{Leese,Ghosh,Mukherjee,Casana, Lima, Lima1,LA}. In the Skyrme model \cite{Adam,Adam1}, a gauge symmetry is required for the scale invariance of a $O(3)$-sigma model with the dynamics of the gauge field governed by Maxwell term \cite{Lima,Casana1}.

It is interesting to mention that there are several types of topological structures. Among these defects, we have the vortices that arise in $3D$ models \cite{Rajaraman,Manton}. Particularly, the study of the vortex solution is interesting because can eventually give us explanations of  some applications, such as the fractional quantum Hall effect \cite{WILCZEK1} and superconductivity at high temperatures \cite{Blaugher}.

To study vortices in our model, we use a technique known as the Bogomol'nyi-Prasad-Sommerfield (BPS) method. This method consists of a series of inequalities for the solutions of partial differential equations that depend on the homotopy class of the solutions at spatial infinity \cite{B,PS}. In other words, the BPS method represents a limit of energy saturation for a given model that allows reducing the order of the equations of motion. The equations of motion obtained at the limit of energy saturation are called BPS equations. 

In this work, we propose an Abelian $O(3)$-sigma model, governed by a Maxwell term, with a Gausson potential, and show that this model admits topological vortex solutions. Gausson models were initially discussed in the context of relativistic theory \cite{Rosen,Rosen1} and non-relativistic one \cite{BBM}. This discussion has recently returned with the study of kinks generated by Gausson-like potentials in ref. \cite{Belendryasova} and with generalized models using Gausson-like logarithmic terms in Refs. \cite{Lima2,Lima3}. It is worth noting that Gausson-like logarithmic potentials are interesting for describing quantum Bose liquids \cite{DZHU}, and quantum gravity theory \cite{ZLOSH}. 

Configurational Entropy (CE) of BPS vortices in an Abelian model with logarithmic interaction is discussed. The term configurational entropy appeared in Refs. \cite{GS,Gleiser3}, underpinned by  information theory of Claude E. Shannon \cite{Shannon}. From a quantum mechanical point of view, Shannon's entropy gives us information on the probability of a particle transitioning from one quantum state to another \cite{Pathria,LRC}. In other words, Shannon's entropy is the hidden information in a random process. The CE was reintroduced as a proposed measure of stability in localized structures. There are several studies using configurational entropy in different systems. For instance, the CE can be used to investigate stable Q-ball solutions \cite{Gleiser14,Gleiser41} at the Chandrasekhar limit for white dwarfs; in the study of the non-equilibrium dynamics of spontaneous symmetry breaking \cite{Gleiser41}; in the study of Bose-Einstein condensates \cite{Casadio}, and in braneworlds to investigate field configurations of multi-kink type \cite{Wilami}.

Our work is organized as follows: In Sec. II, we investigate the Maxwell-Gausson $O(3)$-sigma model through the BPS method. The spherically symmetrical ansatz is considered to demonstrate that magnetic flux is quantized in each topological sector. Using the interpolation method, the topological vortex solutions of the model are obtained numerically. In Sec. III,  we investigate the DCC to analyze the phase transitions that can create new topological structures in the model. Finally, in Sec. IV we discuss our findings.


\section{The Maxwell-Gausson $O(3)$-sigma model}

Motivated by the models presented in Refs. \cite{Mukherjee,Ghosh,Lima1}, we consider the Lagrangian of the Maxwell-Gausson $O(3)$-sigma model in $(2+1)D$ spacetime as
\begin{align}
    \mathcal{L}=\frac{1}{2}D_{\mu}\Phi\cdot D^{\mu}\Phi-\frac{1}{4}F_{\mu\nu}F^{\mu\nu}-\rho\phi_{3}^{2}\text{ln}\bigg(\frac{\phi_{3}^{2}}{\vartheta^{2}}\bigg).
\end{align}

Here $\Phi$ is a triplet of scalar fields, a vector in the internal space that respects the following constraint
\begin{align}
\label{constraint}
    \Phi\cdot\Phi=M^2,
\end{align}
with
\begin{align}
    \phi_i=\Phi\cdot\hat{n}_i.
\end{align}
where we consider $M=1$ (see Ref. \cite{Schroers}), and $\{\hat{n}_i\}$ are a basis of unit orthogonal vectors. The Lagrangian of the model is invariant under an rotation around a preferred axis $\hat{n}_3=(0,0,1)$. 

Note that the component of the $\phi_3$ field is responsible for the spontaneous breaking of symmetry. We use a metric with signature $(+,-,-)$. The $SO(2)$ [$U(1)$] subgroup is gauged by the vector potential $A_\mu$ whose dynamics are dictated by the Maxwell term. The electromagnetic field tensor is
\begin{align}
F_{\mu\nu}=\partial_{\mu}A_{\nu}-\partial_{\nu}A_{\mu}.
\end{align}

The covariant is given by
\begin{align}
    D_{\mu}\Phi=\partial_{\mu}\Phi+A_{\mu}\hat{n}_3\times\Phi,
\end{align}
in natural units, i. e., $\hbar=c=e=1$.

The equations of motion of the model are 
\begin{align}
\label{motion}
    D_{\nu}D^{\nu}\Phi=-\frac{\partial \mathcal{V}}{\partial\Phi},
\end{align}
and 
\begin{align}\label{motion5}
    \partial_{\nu}F^{\nu\mu}=j^\mu,
\end{align}
where $\mathcal{V}=\rho\phi_{3}^{2}\ln(\phi_{3}^{2}/\vartheta^2)$ and $j^\mu=-\textbf{J}^\mu\cdot\hat{n}_3$ with
\begin{align}
    {\bf J}^\mu=-\Phi\times D^\mu\Phi.
\end{align}

\begin{figure}[ht!]
    \centering
    \includegraphics[scale=0.4]{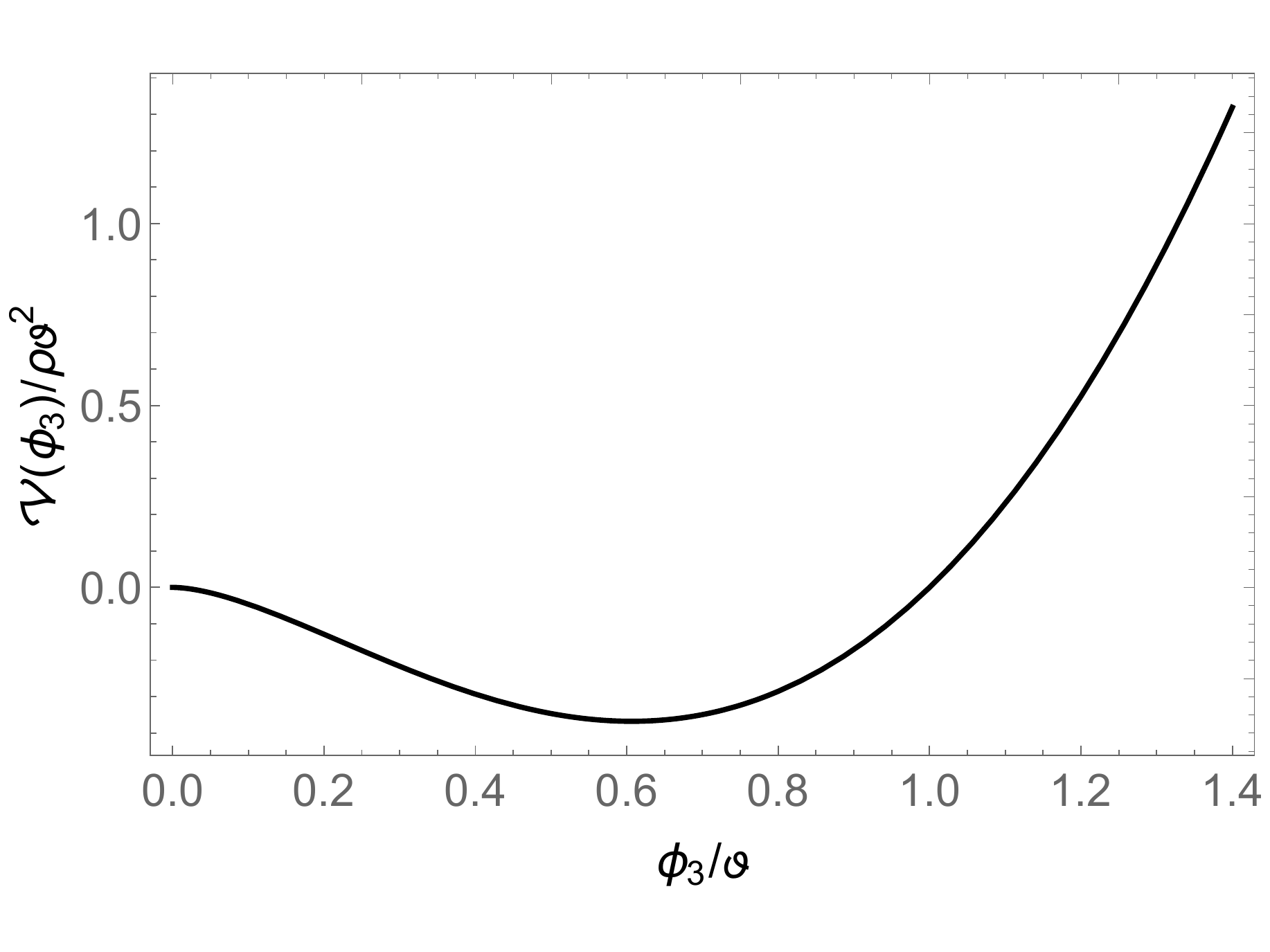}
    \vspace{-20pt}
    \caption{Potential behavior of the model}
    \label{fig1}
\end{figure}

The behavior of the potential responsible for the spontaneous breaking of symmetry is displayed in Fig. \ref{fig1}.

Looking at Gauss' law (component $\mu=0$ in Eq. (\ref{motion5})) and seeking to investigate purely magnetic and stationary vortices (i. e. structures that have a magnetic flux and are independent of time) we choose the gauge $A_0=0$. This leads us to electrically neutral vortex configurations. It is possible to have other vortex solutions that describe more complex structures with electric and magnetic field flux. However, in this work we turn our attention to the study of purely magnetic vortices.

The energy functional is obtained by integrating  $T_{00}$ of the energy-momentum tensor over all space. In this way, the energy functional is
\begin{align}
    \label{energy}
    E=\frac{1}{2}\int\, d^{2}x \bigg[(D_{i}\Phi)^2+(F_{12})^{2}+2\rho\phi_{3}^{2}\text{ln}\bigg(\frac{\phi_{3}^{2}}{\vartheta^{2}}\bigg)\bigg].
\end{align}

The functional that describes the energy of the vortices is rearranged as
\begin{align}
\label{energy1}
    E=&\frac{1}{2}\int\, d^{2}x\bigg\{(D_{i}\Phi\pm\varepsilon_{ij}\Phi\times D_j\Phi)^{2}+\bigg[F_{12}\pm\sqrt{2\rho\phi_{3}^{2}\text{ln}\bigg(\frac{\phi_{3}^{2}}{\vartheta^2}}\bigg)\bigg]^2\bigg\}\pm4\pi\int\, dx^2 \mathcal{Q}_0.
\end{align}

The topological charge of the Maxwell-Gausson O(3)-sigma model \cite{Schroers,Ghosh} is defined as
\begin{align}
    \mathcal{Q}_{\mu}=\frac{1}{8\pi}\varepsilon_{\mu\nu\lambda}\bigg[\Phi\cdot (D^{\nu}\Phi\times D^{\lambda}\Phi)-F^{\nu\lambda}\sqrt{2\rho\phi_{3}^{2}\text{ln}\bigg(\frac{\phi_{3}^{2}}{\vartheta^2}\bigg)}\bigg].
\end{align}

From Eq. (\ref{energy1}) the Bogomol'nyi energy or BPS energy of the model is given by 
\begin{align}
\label{E_BPS}
    E_{BPS}=4\pi\int\, d^2x\mathcal{Q}_0.
\end{align}

The energy is limited by $E_{BPS}$, i. e., 
\begin{align}
    E\geq E_{BPS}.
\end{align}

When $E= E_{BPS}$ the equations that describe the BPS vortex structures of the model are 
\begin{align}
\label{BPS}
    D_i \Phi=\mp\varepsilon_{ij}\Phi\times D_j \Phi,
\end{align}
and
\begin{align}
\label{BPS1}
    F_{12}=\pm\phi_3\sqrt{2\rho\text{ln}\bigg(\frac{\phi_{3}^{2}}{\vartheta^{2}}\bigg)}.
\end{align}

Eqs. (\ref{BPS}) and (\ref{BPS1}) are equations of Bogomol'nyi for the sigma model and similar of those in Ref. \cite{Schroers}.

To investigate the spherically symmetrical vortex structures we assume the well-known ansatz \cite{Schroers,Ghosh,Mukherjee}:
\begin{align}
    \label{ansatz}
    &\phi_1=\sin f(r)\cos n\theta;\\
    &\phi_2=\sin f(r)\sin n\theta;\\
    &\phi_3=\cos f(r).
\end{align}

This ansatz is conveniently chosen so that the $O(3)$-sigma constraint is respected, i.e. 
\begin{align}\nonumber
    \Phi\cdot\Phi=&[\sin^2f(r)\cos^{2} n\theta+\sin^2f(r)\sin^{2} n\theta+\cos^2 f(r)]\\
    =&[\sin^2 f(r)+\cos^2 f(r)]=1.
\end{align}

We consider the behavior of the gauge field \cite{Schroers} is
\begin{align}
    \label{ansatz1}
    \textbf{A}=-\frac{na(r)}{r}\hat{\text{e}}_{\theta},
\end{align}
where $n$ is the winding number of the topological structure \cite{Lima,Lima1}. 

To obtain topological vortex structures, the scalar field and the gauge field must behave as
\begin{align}
    \label{boundary}
    f(r\to 0)=0, \hspace{1cm} f(r\to \infty)=\pi,
\end{align}
and, 
\begin{align}
    \label{boundary1}
    a(r\to 0)=0, \hspace{1cm} a(r\to\infty)=-\eta_1,
\end{align}
which we will discuss in what follows.

Using (\ref{boundary}) and (\ref{boundary1}), the magnetic field of the vortex is
\begin{align}
\label{B}
    \textbf{B}=\nabla\times \textbf{A}\rightarrow B=||\textbf{B}||=-\frac{na'(r)}{r}=-F_{12},
\end{align}
therefore, the magnetic flux $\Phi_B$ of the vortex is 
\begin{align}\nonumber
\label{MagneticField}
    \Phi_{B}=&\int_{S}\int \textbf{B}\cdot d\textbf{S}=-n\int_{0}^{2\pi}\int_{0}^\infty\,\frac{a'(r)}{r}rdrd\theta\\
    \Phi_{B}=&-2\pi n[a(\infty)-a(0)]\rightarrow \Phi_{B}=2\pi n\eta_1,
\end{align}
from the previous expression, it is clear that the magnetic flux of the vortex is quantized in each topological sector.

\subsection{Asymptotic behavior and numerical solution} 

We turn our attention to the study of the asymptotic behavior and numerical solutions describing the magnetic vortex structures. The BPS equations (\ref{BPS}) and (\ref{BPS1}) are written in terms of the variable field, namely,
\begin{align}
    \label{BPS3}
    f'(r)=\pm n\frac{a(r)+1}{r}\sin f(r),
\end{align}
and
\begin{align}
    \label{BPS4}
    a'(r)=\pm 2\frac{r}{n}\cos f(r)\sqrt{\rho\text{ln}\bigg(\frac{\cos f(r)}{\vartheta}\bigg)}.
\end{align}

Let us discuss the asymptotic behavior of the possible solutions of Eqs. (\ref{BPS3}) and (\ref{BPS4}). To ensure no singularity at the origin, the field near the origin must have the form:
\begin{align}
    f(r)=n\pi+\mathcal{O}(r), 
\end{align}
where $n\in \mathbb{N}$.

For this behavior of the variable field $f(r)$ near to the origin, the gauge field must assume a null behavior, i. e. $a(0)=0$. 

To analyze the existence of vortex solutions, we investigate the existence of solutions near to the asymptotic points assumed in Eqs. (\ref{boundary}) and (\ref{boundary1}). First let us consider the initial condition $f(0) = 0$, it is convenient introduce $\chi (r)$ so that $f(r) =\pi+\chi(r)$. For positive winding numbers, the negative sign of Eq. (\ref{BPS3}) is chosen. We consider $\chi (r)\ll 1$ so that the origin the field takes the form
\begin{align}
    \chi(r)=\overline{A}_{0}r^n,
\end{align}
and the gauge field behaves as 
\begin{align}
    a(r)\simeq\frac{1}{2n\sqrt{\rho\text{ln}(\vartheta^{-2})}}\bigg\{-\rho r^2\text{ln}(\vartheta^{-2})+\frac{\overline{A}_{0}^{2}r^{2(n+1)}}{2(n+1)}[\text{ln}(\vartheta^{-2})+1]\bigg\}+\mathcal{O}(r),
\end{align}

On the other hand, for negative winding number $f(r)$ behaves with
\begin{align}
    \chi(r)=\overline{B}_{0}r^{-n},
\end{align}
and
\begin{align}
a(r)\simeq\frac{1}{2n\sqrt{\rho\text{ln}(\vartheta^{-2})}} \bigg\{\rho r^{2}\text{ln}(\vartheta^{-2})-\frac{\overline{B}_{0}r^{2(1+n)}}{2(1+n)}[1+\text{ln}(\vartheta^{-2})]\bigg\}+\mathcal{O}(r),
\end{align}
where $\mathcal{O}(r)$ are the higher order terms. 

Meanwhile, at infinity, the variable field obeys $f(\infty)=\pi$. Similar to the above analysis $f(r)=\pi+\chi(r)$, so that for finite energy configurations $a(\infty)=\xi_1$. In this case,
\begin{align}
    \chi(r)=\overline{C}_{\infty}r^{n(1-\xi_1)}.
\end{align}
and
\begin{align}
    a(r)\simeq \frac{1}{2n\sqrt{\rho\text{ln}(\vartheta^{-2})}}\bigg\{\rho r^2\text{ln}(\vartheta^{-2})-\frac{\overline{C}_{\infty}\rho r^{2[1+n(1-\xi_1)]}}{2-2n(\xi_1-1)}[1+\text{ln}(\vartheta^{-2})]\bigg\}+\mathcal{O}(r).
\end{align}

Finally, for
\begin{align}
    f(\infty)=\pi, \hspace{1cm} \text{and} \hspace{1cm} a(\infty)=\xi_2,
\end{align}
with $n<0$,
\begin{align}
    f(r)=\overline{D}_{\infty}r^{n(1+\xi_2)},
\end{align}
and
\begin{align}
     a(r)\simeq \frac{1}{2n\sqrt{\rho\text{ln}(\vartheta^{-2})}}\bigg\{-\rho r^2\text{ln}(\vartheta^{-2})+\frac{\overline{D}_{\infty}\rho r^{2[1+n(1+\xi_2)]}}{2-2n(1+\xi_2)}[1+\text{ln}(\vartheta^{-2})]\bigg\}+\mathcal{O}(r).
\end{align}

The $\xi_{1,2}$ parameters indicate whether the solutions are topological or non-topological.

\begin{figure}[ht!]
    \centering
    \includegraphics[scale=0.4]{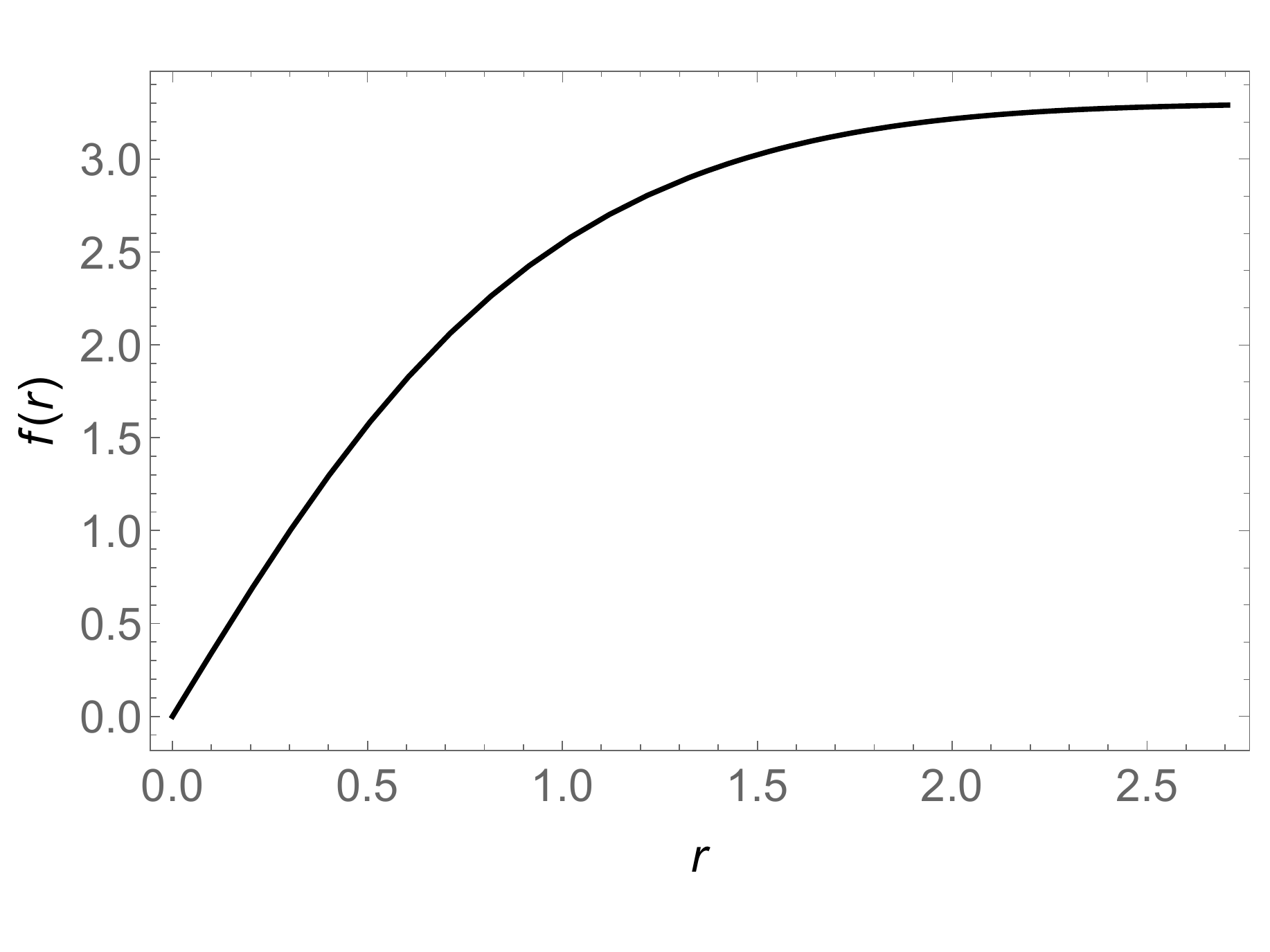}
    \vspace{-1cm}
    \caption{Numerical solution for the variable field $f(r)$. We assume that $n=\rho=1$ and $\vartheta=0.5$.}
    \label{fig2}
\end{figure}

We investigate the numerical solution of the BPS equations (\ref{BPS3}) and (\ref{BPS4}), using the interpolation method (for more details on the interpolation method for solving differential equations see Ref. \cite{Burden}). The result of the variable field of the scalar field is shown in Fig. \ref{fig2}.

\begin{figure}[ht!]
    \centering
    \includegraphics[scale=0.4]{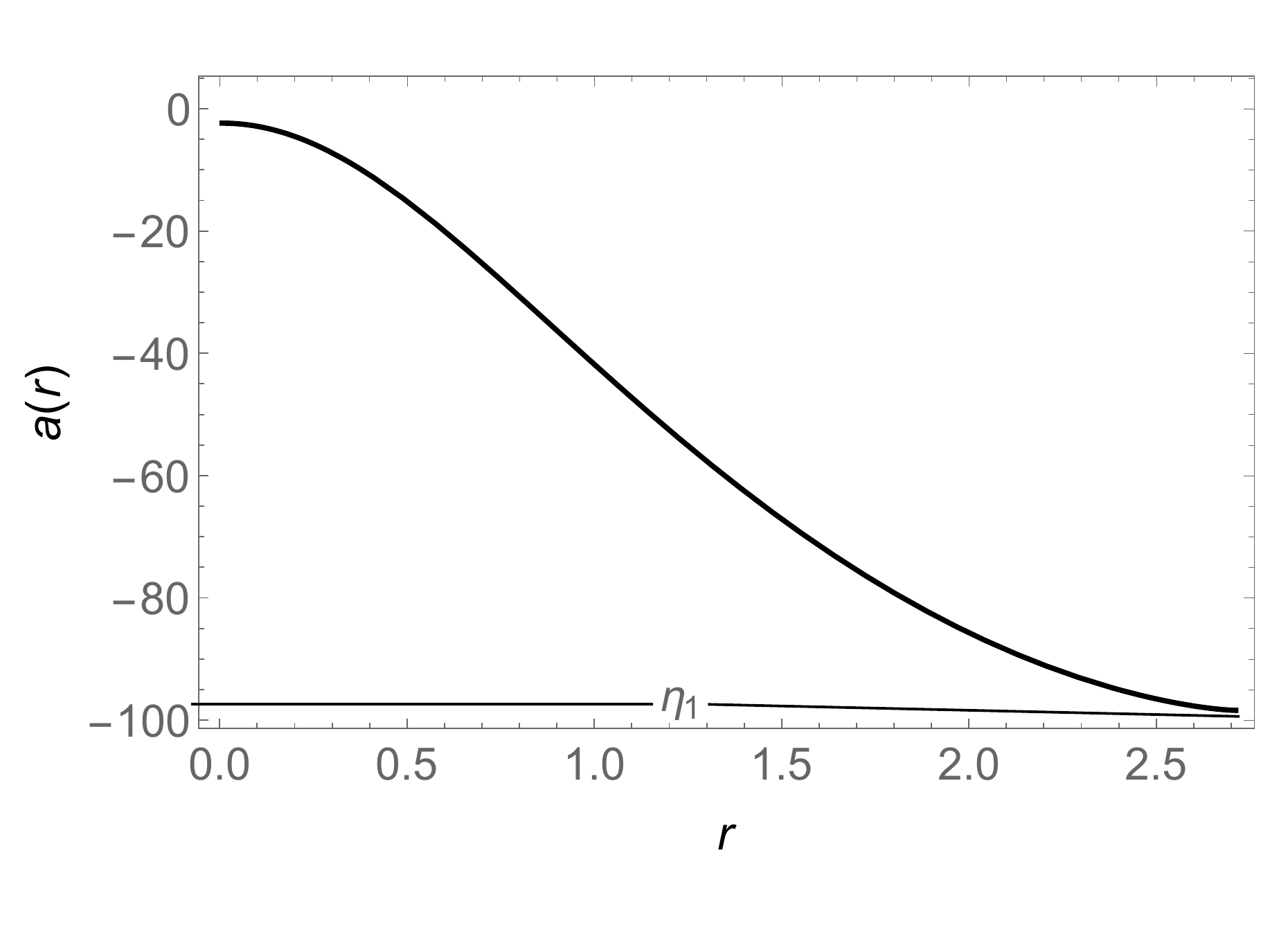}
    \vspace{-1cm}
    \caption{Solution of the field variable corresponding to the gauge field when $n=\rho=1$ and $\vartheta=0.5$.}
    \label{fig3}
\end{figure}

Returning the equations (\ref{BPS3}) and (\ref{BPS4}) and considering the topological boundary conditions (\ref{boundary1}), the solution of the variable field corresponding to the gauge field is obtained and presented in Fig. \ref{fig3}.

\begin{figure}[ht!]
    \centering
    \includegraphics[height=4.4cm]{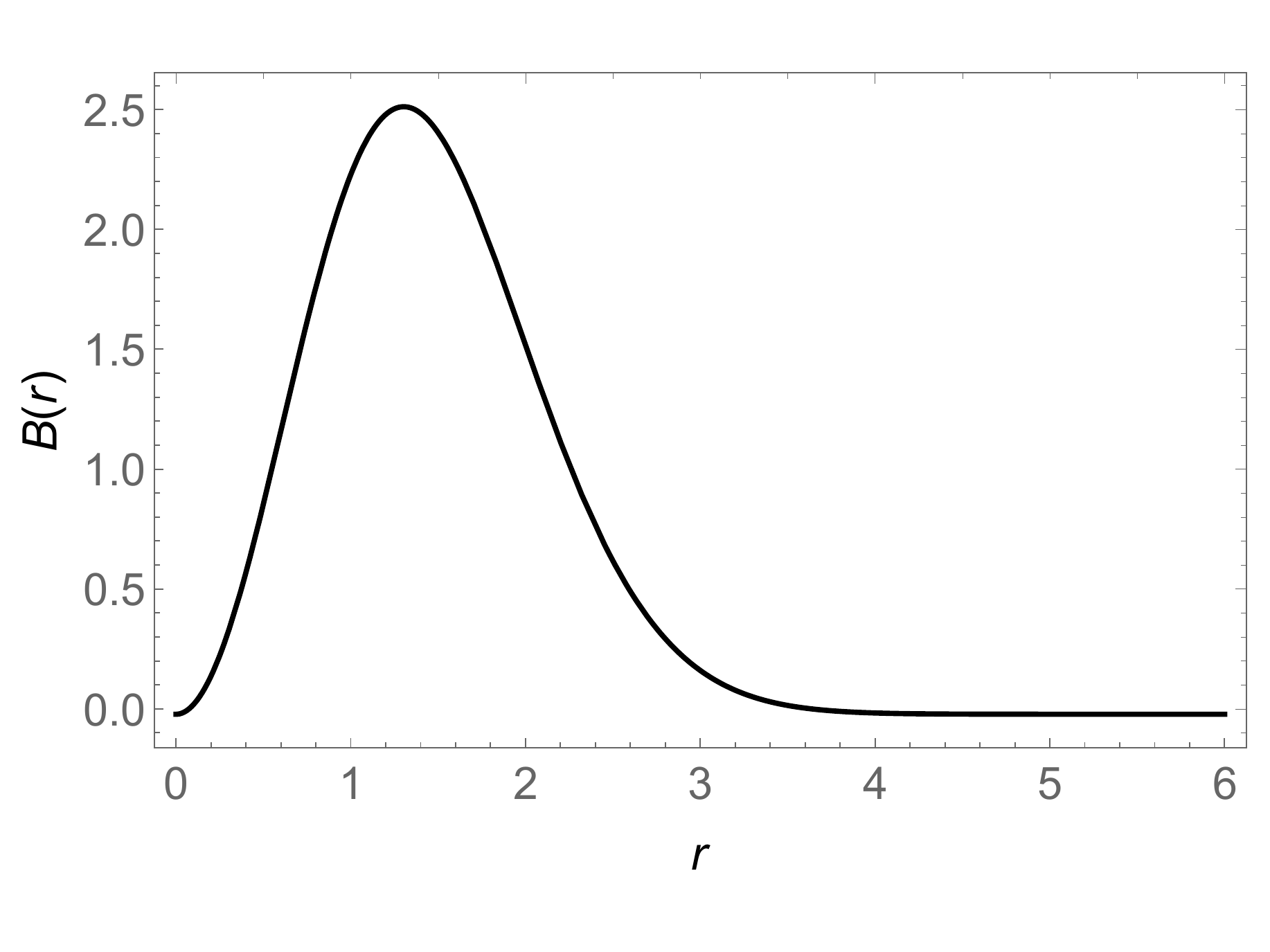}
    \includegraphics[height=4cm]{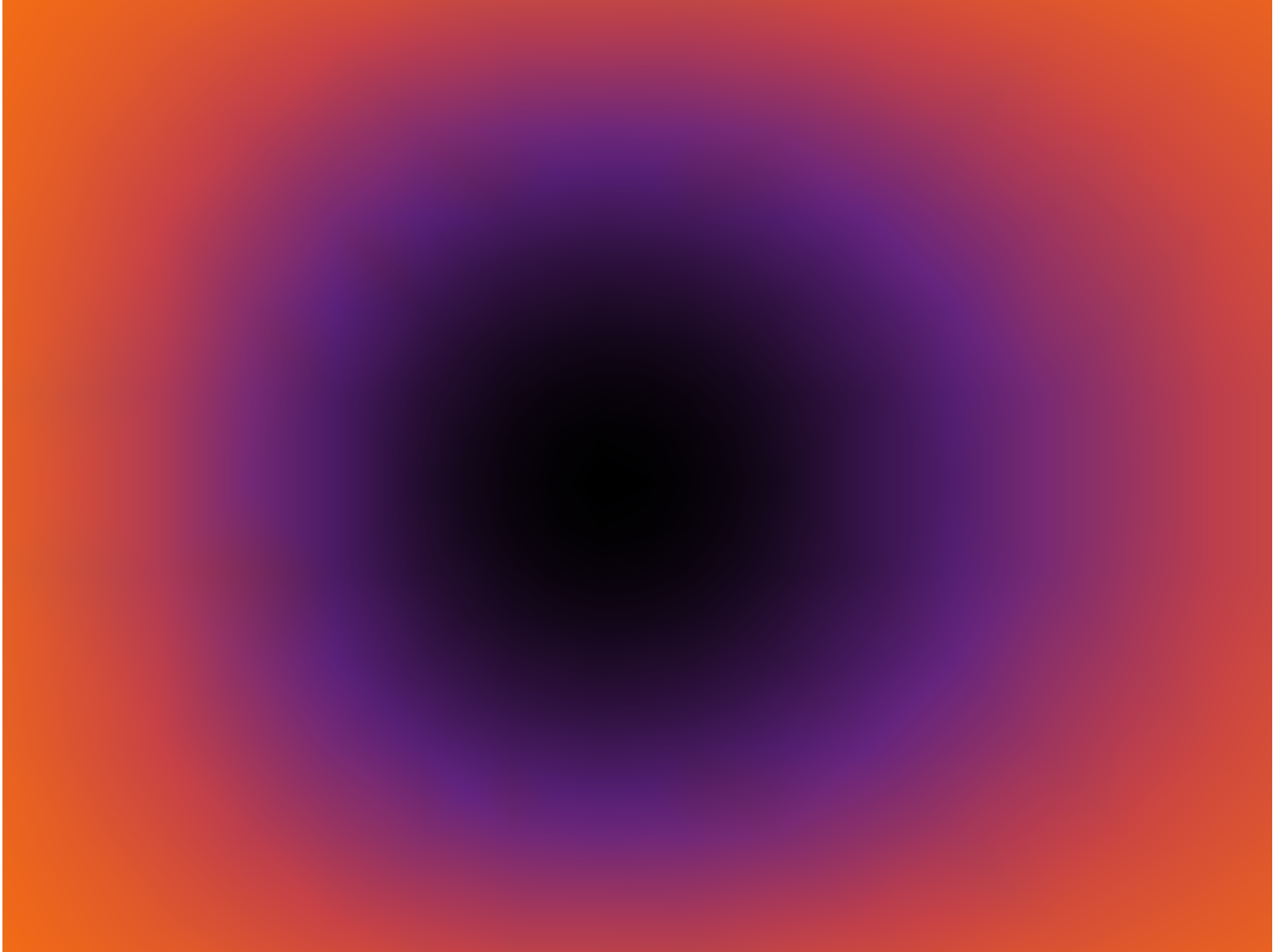}\\
    (a) \hspace{6cm} (b)
    \caption{(a) Magnetic field of the BPS vortex with $n=\rho=1$ and $\vartheta=0.5$. (b) The planar magnetic field of the vortex.}
    \label{fig4}
\end{figure}

From the Eq. (\ref{B}), the numerical solution of the magnetic field responsible for the flux of the vortex investigated. The magnetic field is shown in Fig. \ref{fig4}(a), the planar magnetic field of the vortex is shown in Fig. \ref{fig4}(b).

\begin{figure}[ht!]
    \centering
    \includegraphics[scale=0.4]{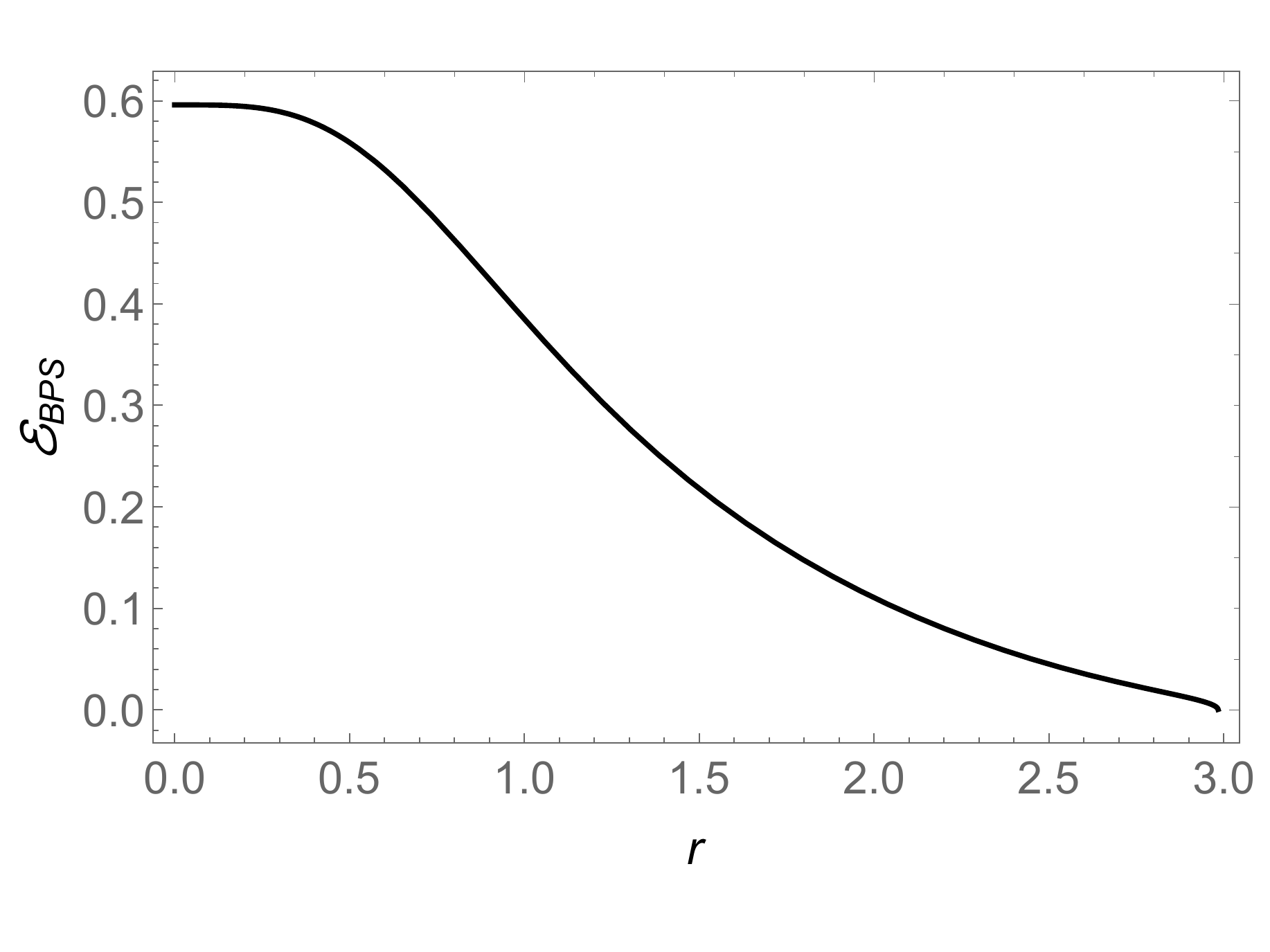}
    \vspace{-1cm}
    \caption{BPS energy density of the vortex with $n=\rho=1$ and $\vartheta=0.5$.}
    \label{fig6}
\end{figure}

The BPS energy of the model can be found from Eq. (\ref{E_BPS}). The corresponding numerical solution is shown in Fig. \ref{fig6}.

From the results found, we observed that Maxwell vortices are purely magnetic due to the fact that we assume $A_0=0$.  From the numerical result of the energy,  it is observed that an intense BPS energy density at the origin. An interesting result arises when the magnetic field is investigated, i. e., due to the $B(r)$ function profile, the planar vortex has a ring-like profile. This behavior of the magnetic field seems to occur due to the fact that the gauge field presents an exponential-like behavior until it reaches the value $-\eta_1$. We emphasize that the shape of solutions for the magnetic field has already been noted in other models \cite{Lima2}

\section{Phase transitions and configurational entropy}

Configurational entropy is inspired by Shannon entropy which was proposed in 1948 \cite{Shannon}. Information in Shannon's communication theory is defined as
\begin{align}\label{S}
    S=-\sum_{i}\rho_i \text{ln} \rho_i,
\end{align}
where $\rho_i$ consists of the probability of message $i$, given the corpus of all possible messages. Shannon's entropy is the minimum number of bits needed to encode a message to achieve a maximal transmission rate of information between a sender and receiver, i. e., the capacity of the channel \cite{Shannon,Gleiser}. In recent years this quantity has been studied in several physical scenarios, such as in the study of quantum mechanical systems \cite{Dong,Dong1,LMMA}.

Motivated by the theory presented by Shannon, configurational entropy emerges. Initially, the configurational entropy appears as the informational complexity of localized field configurations. The messages in configurational entropy are the components of the power spectrum (see for example the Refs. \cite{Gleiser14,Gleiser,Gleiser2}).

For the study of the phase transitions of the BPS topological vortices, we use the concept of differential entropy defined in Refs. \cite{Gleiser3,Gleiser14,Gleiser,Gleiser2,Gleiser4}. The differential entropy is 
\begin{align}
    \mathcal{S}=\int \rho(\textbf{k})\text{ln}\rho(\textbf{k})\, d\textbf{k}.
\end{align}

Although differential entropy is finite, it is not invariant under a change of coordinates. This is a consequence of the fact that the probability density transforms as a scalar density under coordinate transformations $x\to\tilde{x}$ and hence the density transforms as $\rho(x)\to\vert\frac{ \partial\tilde{x}}{\partial x}\vert\rho(\tilde{x})$.

Differential configurational entropy is intended to measure the
formational complexity of a particular field configuration. Consider an energy density $\mathcal{E}(\textbf{r})$ of the field, located in space. The decomposition into wave modes in d-dimensional space is provided by the Fourier transform:
\begin{align}
    \mathcal{G}(\textbf{k})=(2\pi)^{-(d/2)}\int \mathcal{E}(\textbf{r})\text{e}^{-i \textbf{k}\cdot\textbf{r}}\, d^d \textbf{r}.
\end{align}

A detector sensitive to the full spectrum of wave modes will detect a wave mode within of a volume $d^d \textbf{k}$ centered on $\textbf{k}$ with probability proportional to the power in that modes, i. e.,
\begin{align}
    p(\textbf{k}\, \vert d^d \textbf{k})\propto \vert\mathcal{G}(\textbf{k})\vert^2\, d^d\textbf{k}.
\end{align}

The relative contribution of a wave mode to a $\vert\textbf{k}_{*}\vert$ is the modal fraction, 
\begin{align}
g(\textbf{k})=\frac{\vert\tilde{\mathcal{G}}(\textbf{k})\vert^2}{\vert\tilde{\mathcal{G}}(\textbf{k}_{*})\vert^2}.
\end{align}

The DCC is 
\begin{align}
    \mathcal{S}_{\mathcal{C}}=-\int g(\textbf{k})\text{ln} [g(\textbf{k})]\, d^d \textbf{k},
\end{align}
since the modal fraction is $\leq 1$, the DCC is positive definite.

For the spherical symmetry, the hyperspheric Fourier transform is given by
\begin{align}
    \mathcal{G}(k)=k^{1-\frac{d}{2}}\int_{0}^{\infty}\mathcal{E}(r)J_{\frac{d}{2}-1}(k r)\, r^{d/2}\,dr
\end{align}
where $J_\nu$ is the Bessel function and $\mathcal{E}(r)$ is the energy density of the vortex. 

The differential configurational complexity is calculated as 
\begin{align}\label{DCES}
    \mathcal{S}_{\mathcal{C}}=-\frac{2\pi^{d/2}}{\Gamma(\frac{d}{2})}\int_{0}^{\infty}k^{d-1}g(k)\text{ln} [g(k)]\,dk,
\end{align}
where the entropy density $\rho_{E}$ is the DCC integrand. 

It is important to mention that in the context of braneworlds, the configurational entropy is able to describe new topological structures, for instance, multi-kink solutions which are fruits of multiples phase transition \cite{Wilami}. In the context in which we are applying the study of this quantity, we want to understand whether the vortex solutions discussed above are unique and whether our nonpolynomial model admits such multiple transitions.

\begin{figure}[ht!]
    \centering
    \includegraphics[scale=0.5]{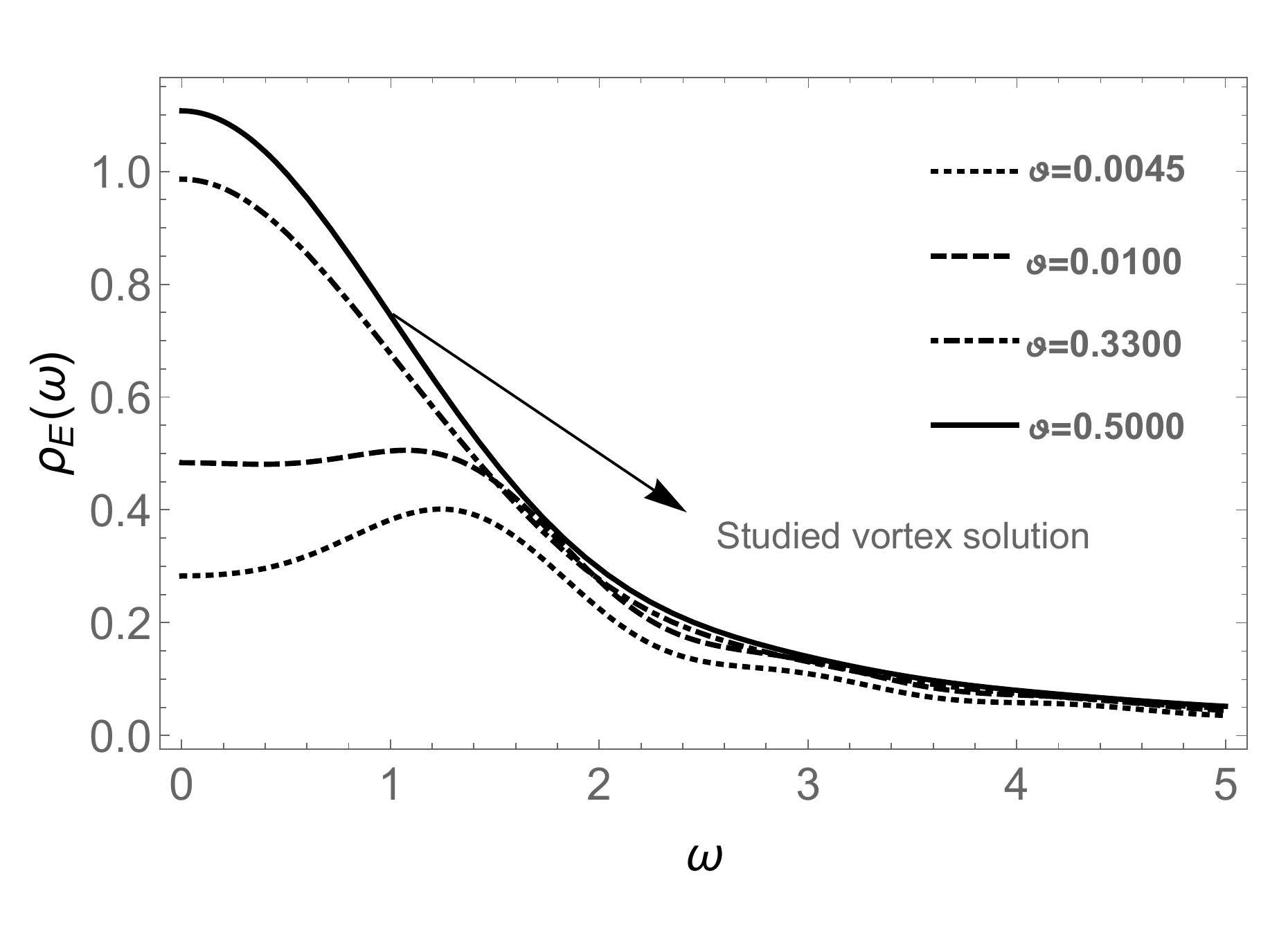}
    \vspace{-1cm}
    \caption{Configurational entropy density for the variable field of the vortex.}
    \label{ed}
\end{figure}

The calculations of the DCC are not so simple, due to our energy density profile in field internal space. Conscious of the difficulties, we evaluate numerically the entropic density of the model (where is spherical symmetry considered) and show the results in Fig. \ref{ed}. To obtain this entropic densities we consider the numerical vortex solutions of Eqs. (\ref{BPS3}) and (\ref{BPS4}) in Eq. (\ref{DCES}). We aware that the parameter $\vartheta$ is responsible for changing the topological structures of the model, since it is related to the vacuum expectation value (v.e.v.). Therefore, we study numerically how this quantity influences the Shannon entropy and how it could generates possible topological structures in the model and consequently new phase transitions. 

\begin{figure}[ht!]
    \centering
    \includegraphics[height=6.5cm]{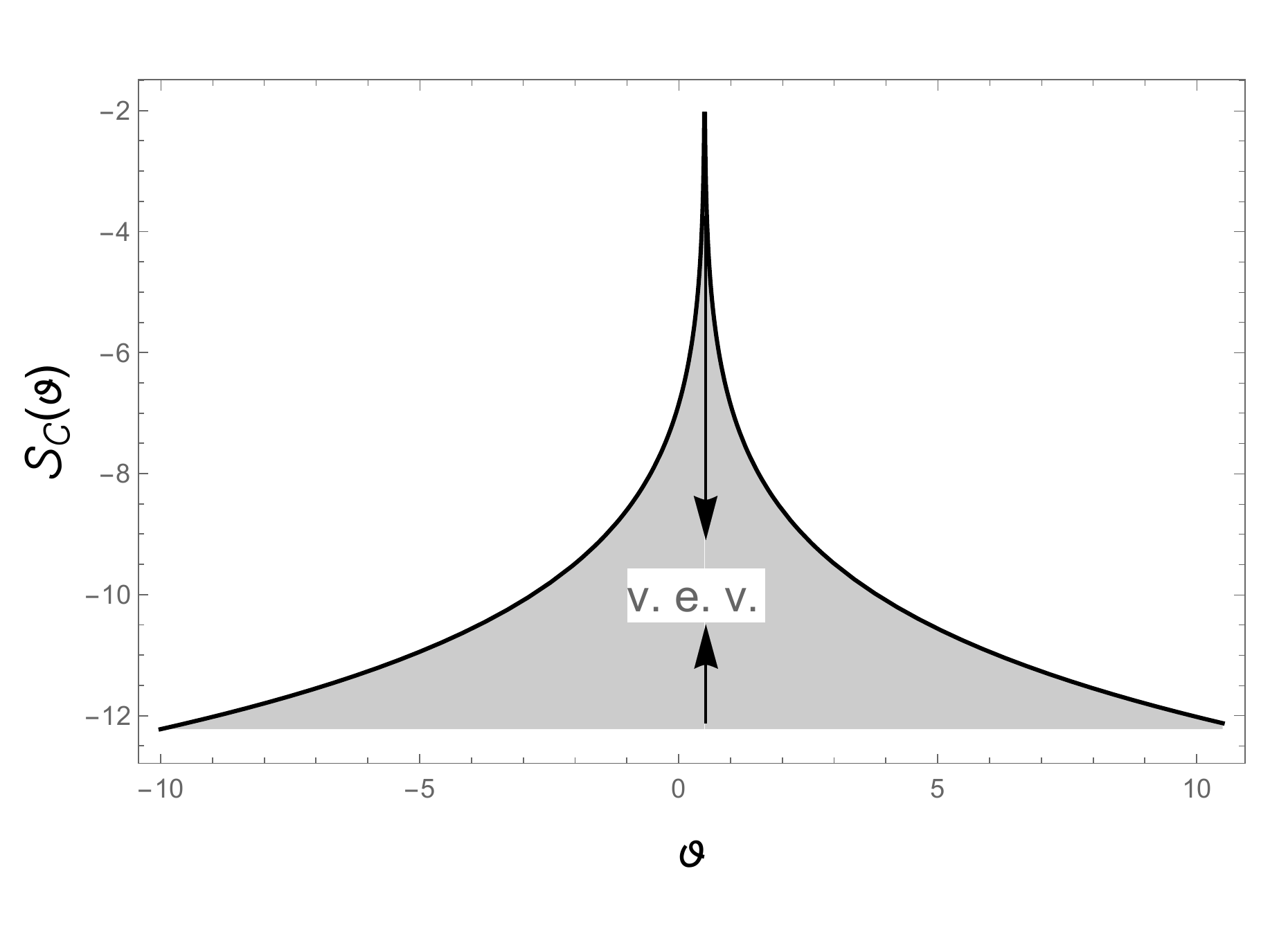}
    \vspace{-1cm}
    \caption{Configurational entropy associated with the variable field of the vortex.}
    \label{ce}
\end{figure}

From the numerical solution of entropic density, is investigated the configurational entropy of the model. The numerical result is shown in Fig. \ref{ce}. In fact, it is noticed that around the core of the vortex, i. e., $r\approx 0$, the entropy densities are greater as the parameter $\vartheta$ increases. In fact, due to the localized structure is around $r=0$ the entropic density is higher than all region. We note that the parameter $\vartheta$ (associated with v.e.v.) that controls the entropic density is associated with the magnetic field of the vortex. So, when this parameter increases, the magnitude of the magnetic field decreases, making the entropic density with a more localized profile. From Fig. (\ref{ce}) for configurational entropy, it is observed that the $O(3)$-sigma model with the Maxwell gauge field and with a nonpolynomial potential type $\phi^4$ admits only kink-like solution. This solution is shown in Fig. (\ref{fig2}). This result, is unlike the results presented in braneworld scenarios, and due to the configurational entropy profile of the $O(3)$-sigma model, our model does not support multiple phase transitions. It is noticed that the critical point of the vortex configurational entropy corresponds to the vacuum expectation value.

\section{Conclusion}

By choosing a self-interaction logarithmic potential (called Gausson potential) with symmetry breaking in the $O(3)$-sigma model gauged with the Maxwell gauge fields, we get purely neutral magnetic vortices. These vortices are topologically stable having quantized magnetic flux $\Phi_B=2\pi\eta_1 n$ in each topological sector. The arising of a magnetic field like a ring in the planar model is observed. This result seems to be a consequence of the behavior of the gauge field in the region. Finally, it is observed that these vortices share the desirable feature of breaking of the scale-invariance.

Considering the spatial profile of the energy density of the model we investigated whether the studied vortices admitted multiple phase transitions. It is observed that the configurational entropy assumed a profile delta-function type centered in  the vacuum expectation value. Due to this profile of the configurational entropy it is clear that the model does not support multiple phase transitions. Meanwhile, by numerical simulation it is concluded that all topological solutions  supported by the model are limited to the range $-10\leq\vartheta\leq 10 $. In fact, in this range of values, the solutions will differ only ``how quickly'' the fields evolve into a vacuum state.

\section*{Acknowledgments}
The authors thank the Conselho Nacional de Desenvolvimento Cient\'{\i}fico e Tecnol\'{o}gico (CNPq), grant n$\textsuperscript{\underline{\scriptsize o}}$ 308638/2015-8 (CASA), and Coordena\c{c}ao de Aperfei\c{c}oamento do Pessoal de N\'{\i}vel Superior (CAPES), for financial support. Also, the authors thank the anonymous referee for his kind and fair criticisms and suggestions.

\end{document}